\documentclass[apl,aps,prb,preprint,groupedaddress,preprintnumbers,showpacs]{revtex4}
\usepackage[T1]{fontenc}
\usepackage{graphicx}% Include figure files
\usepackage{dcolumn}% Align table columns on decimal point

\usepackage{color,graphics,graphicx,fancybox,fancyhdr}
\usepackage{latexsym}
\usepackage{amsmath}
\usepackage{amssymb}

\begin{document}
\bibliographystyle{apsrev}

\graphicspath{../Articolo} \DeclareGraphicsExtensions{.eps,.ps}

\title{Discrete model analysis of the critical current density measurements in superconducting thin films by a single coil inductive method}

\author{M. Aurino$^1$, E. Di Gennaro$^1$, F. Di Iorio$^1$, A. Gauzzi$^2$, G. Lamura$^{1,*}$ and A. Andreone$^1$}

\affiliation{$^1$CNR-I.N.F.M. COHERENTIA and Dipartimento Scienze
Fisiche, Universit\`{a} di Napoli Federico II, I-80125, Napoli,
Italy.\\ $^2$Istituto Materiali per Elettronica e Magnetismo -
Consiglio Nazionale delle Ricerche, Area delle Scienze, 43010
Parma, Italy.}

\date{August 8, 2005}

\begin{abstract}
The critical current density of a superconducting film can be
easily determined by an inductive and contactless method. Driving
a sinusoidal current in a single coil placed in front of a
superconducting sample, a non zero third harmonic voltage $V_{3}$
is induced in it when the sample goes beyond the Bean critical
state. The onset of $V_{3}$ marks the value of current beyond
which the sample response to the magnetic induction is no more
linear. To take into account, in a realistic way, the magnetic
coupling between the film and the coil we have developed a
discrete model of the inducing and induced currents distribution.
In the framework of this model the magnetic field profile on the
sample surface and the coefficient linking the current flowing in
the coil and the critical current density $J_{C}$ of
superconducting thin films is evaluated. The numerical results are
checked measuring $J_{C}$ of several thin films of
$YBa_{2}Cu_{3}O_{_{7-\delta}}$ of known superconducting
properties, used as a control material.

\end{abstract}

%insert suggested PACS numbers in braces on next line

%\maketitle must follow title, authors, abstract and PACS

\maketitle

\section{Introduction}

The use of superconductors in engineering applications like high
field superconducting magnets is of fundamental importance. For
this reason the critical current density is a relevant figure of
merit, that needs to be measured in samples of different nature
and shapes like sintered pellets, tapes or thin films. In
particular, the development of flat cables for superconducting
magnets requires a flexible and reliable measurement of this
transport property without damaging the sample.

Years ago Claassen \textit{et al.} \cite{Classen1} proposed a non
destructive and contactless method to measure J$_{C}$ in
superconducting thin films. This method was theoretically
justified \cite{MAWA1}, subsequently extended for the analysis of
bulk superconductors \cite{MAWA2}, and then used for the
determination of the current-voltage characteristics of
superconducting thin films \cite{Yamasaki}. The basic principle is
the following: a small coil, placed in close contact with the
sample under test, is excited injecting a sinusoidal current
$I_{0}cos\omega t$, and an ac magnetic field is generated. The
linear and nonlinear response of the superconducting film is then
collected by the same coil. The critical current density J$_{C}$
is determined by looking at the behaviour of the induced third
harmonic signal $V_{3}cos(3\omega t)$ as a function of the
inducing current. The link between J$_{C}$ and the current
circulating in the probe coil is found\cite{Classen1,MAWA1} by
assuming two important conditions: \textit{i)} the sample surface
is flat and extended to infinity in two dimensions; \textit{ii)}
the total induction field is zero on the back side of the sample.
Of course this last condition follows from the first one when the
magnetic penetration depth $\lambda\leq d$, where $d$ is the
sample thickness. Due to these assumptions, the field profile must
have as boundary condition that only the parallel component is
nonzero at the sample surface. By using a "mirror technique", the
sample is replaced by an image coil reflected by the sample
surface and carrying the same current. In this case the total
induction field on the sample surface is just twice the value that
would be generated by the single coil in the absence of the
screening currents. According to \textit{ii)} and the Ampère's
law, the maximum sheet current $K_{s}$ that can be sustained by a
superconducting film placed at a distance \textit{h} from the coil
results to be
$K_{S}(h)=\frac{1}{\mu_{0}}B_{r}^{TOTAL}(h)=\frac{1}{\mu_{0}}2\cdot
max \{ B_{r}(h) \}$ where $B_{r}(h)=\mu_{0} I_{0}f(r,h)cos(\omega
t) $ is the radial component of the induction field generated by
the coil at distance $h$ in the absence of the sample. Applying
the Bean's model, $K_{S}=J_{c}d$. By equaling the two expressions
found for $K_{S}$, the scaling relation between the critical
current density and the inducing coil current follows:

\begin{eqnarray}\label{0}
J_{c}=k\cdot\frac{I_{C0}}{d}
\end{eqnarray}

where $I_{C0}$ is the inducing critical current (defined, for
example, using a threshold criterion) and $k=2*max \{ f(r,h) \}$
is the coil factor.

In this paper we review this contactless technique by using a
discrete modelisation for the current distribution in the coil, in
order to calculate in an exact way the induction field in all the
space as a function of the sample distance to the coil, thickness
and magnetic penetration depth. We verify how and under which
circumstance the assumptions made are verified in the framework of
the model, and then we calculate the scaling coefficients linking
the current circulating in the coil to the critical current
density of the sample. The paper is organized as follows: in
section II, after a brief overview on previous studies, we present
the model and calculate the vector potential in all the space for
the sample coil configuration; in section  III we check the
validity of our model comparing our analysis with similar results
presented in literature; in section IV we present the experimental
set-up used to measure superconducting samples of known transport
properties to validate the analytical results; in section V we
finally draw some conclusions.

\section{The discrete model}

The Non Destructive Evaluation (NDE) of the conducting properties
of metallic specimens by using eddy currents methods has been
since years an important subject for the scientific community for
both applicative and fundamental issues. As an example, we can
cite the pioneering work by C. P. Bean\cite{BEAN2} that used a
simple solenoid to measure the resistivity of a sample placed
inside. The necessity of studying large area samples raised the
problem of calculating the spatial profile of the induction field
when using a small multiturn coil. This problem was solved in an
original way by Dodd and Deeds\cite{DODD} with an "hybrid" model:
they calculated the field profiles for a single turn coil and
extended the result to the case of multiturn coils by integrating
over the coil section. Their work has been the starting point for
hundreds of experimental and theoretical studies in the field of
NDE.

In the case of superconducting materials, the thin film technology
raised the interest in the study of both the transport
properties\cite{LORENZ} (the critical current density) and the
electrodynamic response (the absolute magnetic penetration
depth)\cite{FIORY,JEANNERET,LIN,LEMBE1,LEE,PRUSSEIT,CLASSEN3,COFFEY0}
without damaging the sample under test. For these issues, a two
coil configuration with both continuous and discrete approach has
been used.

For a single coil configuration, a completely discrete approach by
Gauzzi \textit{et al} \cite{GAUZZI} was used to model the coil
impedance variation for the measure of the magnetic penetration
depth of superconducting samples of different nature and shapes.
Recently M. W. Coffey \cite{COFFEY} has reviewed the single and
multiturn impedance coil problem by using a continuous method of
kernel functions. For the evaluation of the transport properties
of superconducting samples, we have shortly presented in the
introduction the continuous model approach\cite{Classen1,MAWA1}.

In this section, we extend the discrete model by Gauzzi \textit{et
al} \cite{GAUZZI} to the calculation of the field induction
profiles in all the space. We will show how and in which cases
this exact model is simpler and more precise respect to previous
hybrid or continuous models cited above.

\begin{figure}
\textwidth 9 cm
  \centering
  \includegraphics[scale=0.9]{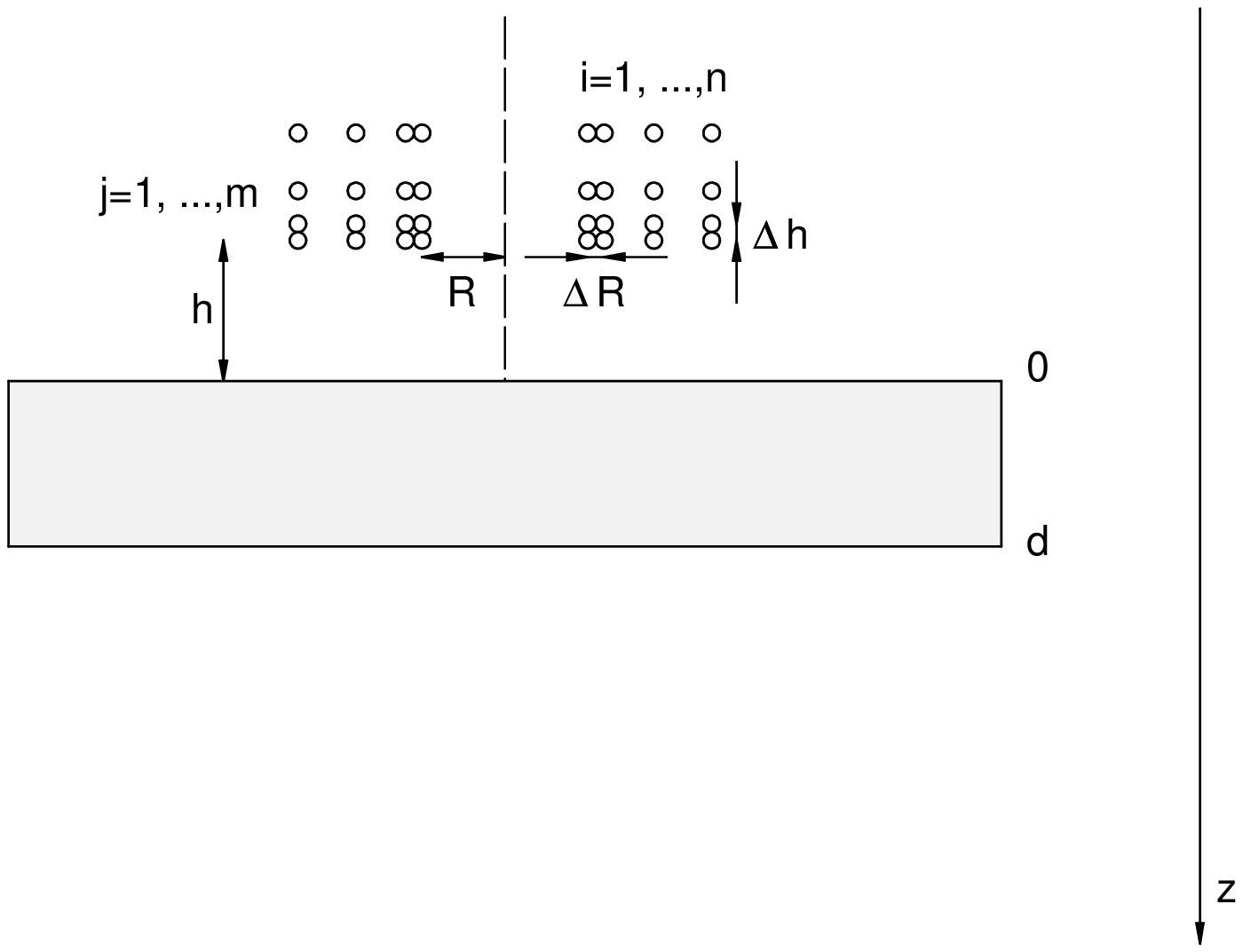}
  \caption{Section of the sample-coil configuration. The coil is
modelled as a discrete distribution of concentric turns axially
and radially equispaced.  $h$ is the sample-coil distance and $d$
is the sample thickness. $\Delta h$ and $\Delta R$ are the
spacings between N adjacent sheets and M adjacent turns in the
same sheet respectively.}
  \label{fig1}
\end{figure}

We investigate the ac response of a superconducting slab of finite
thickness situated at $-d<z<0$, at a distance $h$ from a pancake
coil, as shown in figure ~\ref{fig1}. Further, we consider the
hypothesis that the outer radius of the coil is less than half
sample size to neglect edge effects. This will be better clarified
in the following, when we will evaluate and show the magnetic
field profiles.

First, we model the coil as a discrete distribution of concentric
turns axially and radially\cite{GAUZZI} equispaced. The wire is
supposed infinitely thin; this assumption is valid when the wire
diameter is much smaller than the skin depth at the working
frequency. For example, in the case of a copper wire and at
frequencies below MHz this approximation holds for wire diameter
up to 0.1 mm. Thus the total current density circulating in the
coil can be written in cylindrical coordinates in the following
way:

\begin{eqnarray}\label{1}
J_{\theta,coil}(r,z)=I\sum_{n=0}^{N-1}\delta(z+h+n\Delta
h)\sum_{m=0}^{M-1}\delta(r-R-m\Delta R)
\end{eqnarray}

where I is the total current flowing in the coil, $\delta$ is the
delta function, $R$ is the internal coil radius and $\Delta h$,
$\Delta R$ are the spacings between N adjacent sheets and M
adjacent turns in the same sheet respectively (see fig.1). We
assume that $\Delta h$ and $\Delta R$ both equals the wire
diameter $d_{w}$.

Secondly, we consider the Maxwell equation at a frequency
$\omega/2\pi$ for the vector potential $\overrightarrow{A}$ in the
London gauge ($\overrightarrow{\nabla} \cdot \overrightarrow{A}=
0)$, explicitly separating the contributions to the current
density in the coil and in the sample:

\begin{eqnarray}\label{2}
\nabla^{2}A_{\omega}=-\mu_{0}(j_{\omega,coil}+j_{\omega,sample}).
\end{eqnarray}

At low frequencies within the limit of validity of Ohm's law, the
induced screening current in the sample can be expressed in the
following way:

\begin{eqnarray}\label{3}
j_{\omega,sample}=i\omega\sigma_{\omega}A_{\omega}
\end{eqnarray}

where $\sigma_{\omega}$ is the frequency dependent complex
conductivity of the sample. By inserting eq.~\ref{3} into
eq.~\ref{2}, the solution for the field $A_{\omega}$ becomes a
function of $j_{\omega,coil}$ and $\sigma_{\omega}$, the latter
being the only free parameter.

To solve eq.~\ref{2}, we note that the current density j$_{coil}$
in the coil lies in the (x-y) plane of the sample.The same must be
true for the induced screening currents in the sample, as in a
mirror image. This consideration, and the assumption of a sample
with isotropic conductivity $\sigma$ in the plane, allow us to
express eq.~\ref{2} in cylindrical coordinates. In this case, only
the orthoradial component $A_{\vartheta}$ of the vector potential
$\overrightarrow{A}$ is non zero:

\begin{eqnarray}\label{4}
\nabla^{2}A_{\vartheta}=-\mu_{0}(j_{\omega,coil}+iH_{d}(z)\omega\sigma_{\omega}A_{\vartheta})
\end{eqnarray}

where the origin and direction of the z-axis are chosen as shown
in figure ~\ref{fig1}, the $\omega$-subscripts have been omitted,
and $H_{d}(z)$ is the Heaviside function, whose value is unitary
in the [0,d] interval and zero elsewhere. We have solved the
differential equation ~\ref{4} in all the space with and without
the sample by following the Pearl's formalism \cite{Pearl,GAUZZI}.
The solution for the vector potential without the sample is:

\begin{eqnarray}\label{5}
A_{\vartheta}(r,z)=-\mu_{0}\frac{I_{0}}{2}\int^{\infty}_{0}
\left[\sum_{n} e^{-|z+h_{n}|\gamma} \sum_{m} r_{m}
J_{1}(r_{m}\gamma)\right] J_{1}(r \gamma) d\gamma
\end{eqnarray}

In the presence of a superconducting sample we have three
different solutions, corresponding to three different regions:

z<0

\begin{eqnarray}\label{6}
A_{\vartheta}(r,z)=\mu_{0}\frac{I_{0}}{2}\int^{\infty}_{0}
\left[\sum_{n} e^{-|z+h_{n}|\gamma} \sum_{m} r_{m}
J_{1}(r_{m}\gamma)\right] \cdot \left(1-F(\gamma,z,d,l)\right)
J_{1}(r\gamma) d\gamma
\end{eqnarray}

\begin{eqnarray}\label{7}
F(\gamma,z,d,l)= \frac{ 1} {1+ 2 \gamma^{2}l^{2} +2\gamma l
\sqrt{1+l^{2} \gamma^{2}} \coth \left( \frac{d}{l} \sqrt{1+l^{2}
\gamma^{2}} \right)}
\end{eqnarray}

$0\leq z \leq d$

\begin{eqnarray}\label{8}
A_{\vartheta}(r,z)=-\mu_{0}\frac{I_{0}}{2}\int^{\infty}_{0}
\left[\sum_{n} \exp(-h_{n}\gamma) \sum_{m} r_{m}
J_{1}(r_{m}m\gamma)\right] G(\gamma,z,d,l) J_{1}(r\gamma) d\gamma
\end{eqnarray}

\begin{eqnarray}\label{9}
G(\gamma,z,d,l)=\frac{ 2 \gamma l \left[\gamma l \sinh \left(
\frac{z-d}{l} \sqrt{1+l^{2} \gamma^{2}} \right) - \sqrt{1+l^{2}
\gamma^{2}} \cosh \left(\frac{z-d}{l} \sqrt{1+l^{2} \gamma^{2}}
 \right) \right]} {(1+ 2 \gamma^{2}l^{2}) \sinh \left( \frac{d}{l}
\sqrt{1+l^{2}\gamma^{2}} \right)+ 2 \gamma l \sqrt{1+l^{2}
\gamma^{2}} \cosh \left(\frac{d}{l} \sqrt{1+l^{2} \gamma^{2}}
\right)}
\end{eqnarray}

z>d

\begin{eqnarray}\label{10}
A_{\vartheta}(r,z)=-\mu_{0}\frac{I_{0}}{2}\int^{\infty}_{0} \left[
\sum_{n} e^{-h_{n}\gamma} \sum_{m} r_{m} J_{1}(r_{m} \gamma)
\right] \cdot \left[1-e^{-z \gamma} -L(\gamma,l, d) e^{-d \gamma}
\right] J_{1}(r \gamma) d \gamma
\end{eqnarray}

\begin{eqnarray}\label{11}
L(\gamma, l, d)=\frac{ 2 \gamma l \sqrt{1+l^{2} \gamma^{2}}}{(1+ 2
\gamma^{2}l^{2}) \sinh \left( \frac{d}{l} \sqrt{1+l^{2}\gamma^{2}}
\right)+ 2 \gamma l \sqrt{1+l^{2} \gamma^{2}} \cosh
\left(\frac{d}{l} \sqrt{1+l^{2} \gamma^{2}} \right)}
\end{eqnarray}

where $\gamma$ is a wave number in the xy-plane, the functions
\textit{F}, \textit{G} and \textit{\textit{L}} represent the
sample contribution in the considered region, and $J_{0}$ and
$J_{1}$ are the Bessel functions of order zero and one
respectively. $l$ is a complex length expressed by the formula:

\begin{eqnarray}\label{12}
l=\sqrt{\frac{i}{\mu_{0}\omega \sigma_{\omega}}}.
\end{eqnarray}

Note that it reduces to the normal skin depth for the case of
normal metals and to the magnetic penetration depth for the case
of superconductors. The radial and axial components of the
magnetic field with and without the sample have been calculated
from the vector potential by using the standard recursive
relations:

\begin{eqnarray}\label{13}
B_{r}(r,z)=-\frac{\partial A_{\vartheta} }{\partial z }
\end{eqnarray}

\begin{eqnarray}\label{14}
B_{z}(r,z)=\frac{1}{r}\frac{\partial}{\partial r}(r A_{\vartheta})
\end{eqnarray}

The explicit expressions for the components of the induction field
in all the space are reported in appendix A.

\section{Numerical results}

\begin{figure}
\textwidth 9 cm
  \centering
  \includegraphics[scale=0.8]{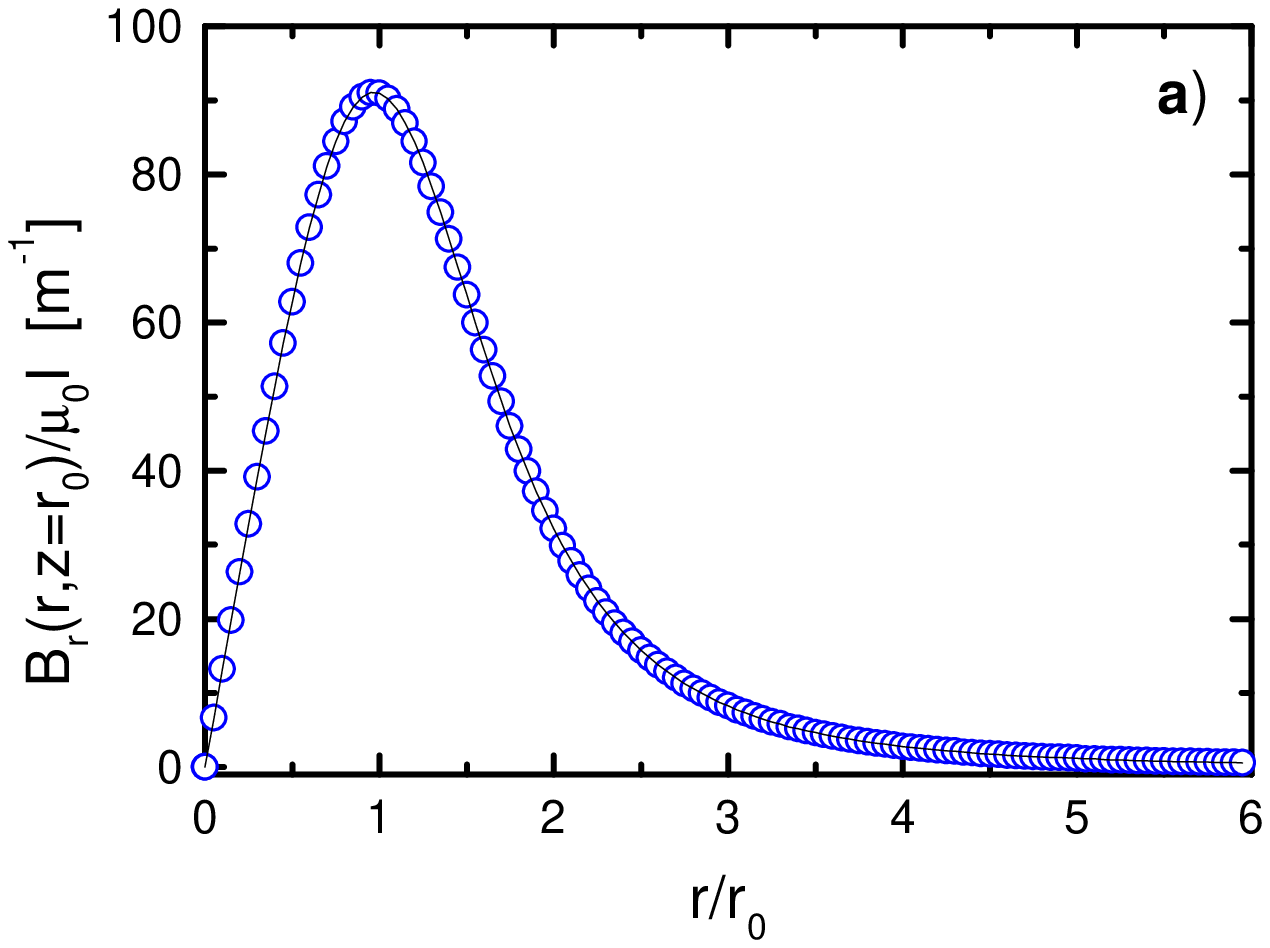}
  \includegraphics[scale=0.8]{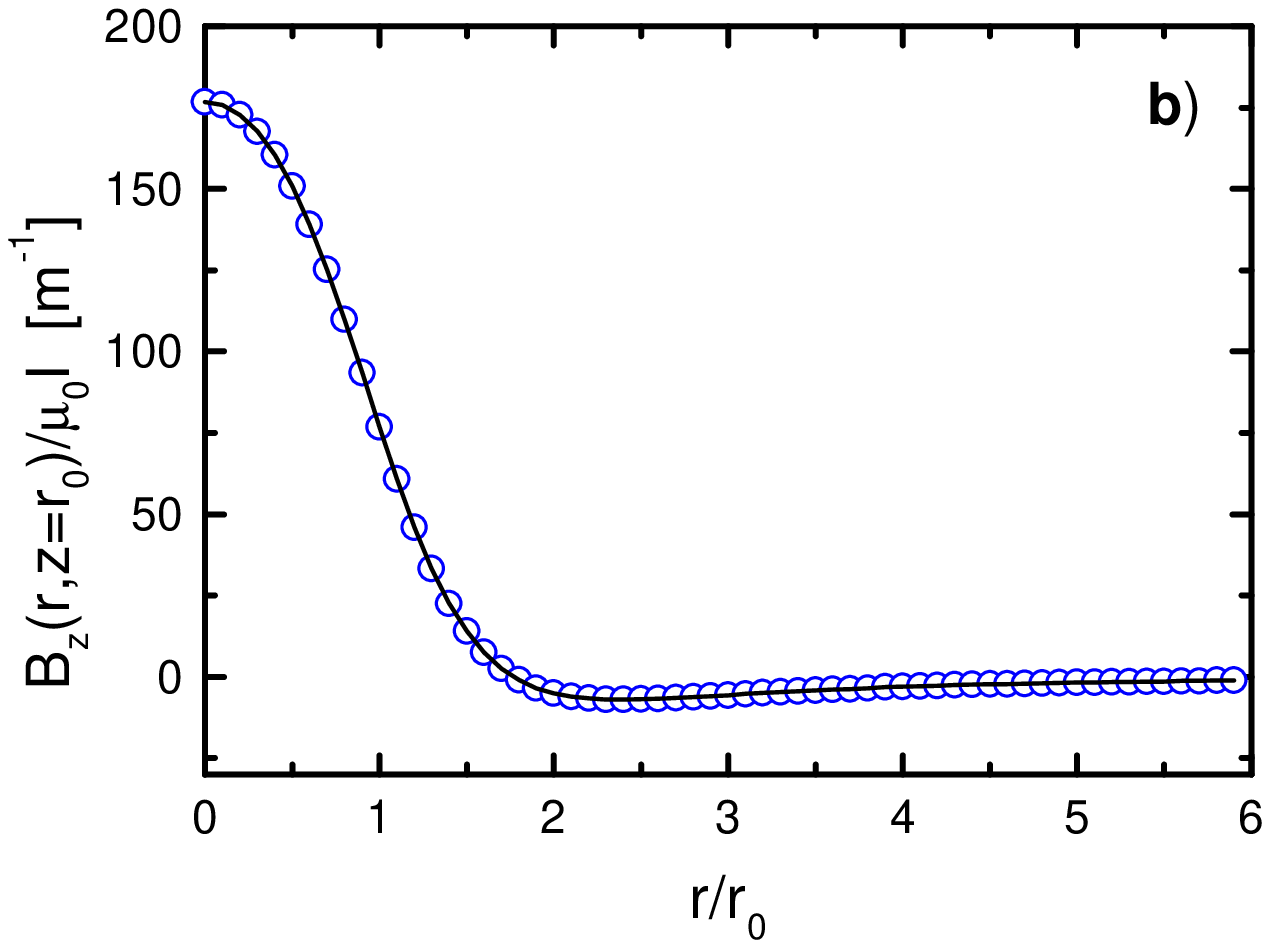}
  \caption{(Color Online) a) radial component $B_{r}(r,z=r_{0})$ of
the induction field for a one turn one sheet coil calculated by
using the Dodd and Deeds's model (continuous line) and our
discrete model (open circles) respectively  as a function of
$r/r_{0}$; b) the same as in a) for the normal component
$B_{z}(r,z=r_{0})$.}
  \label{fig2}
\end{figure}

\begin{figure}
\textwidth 9 cm
  \centering
  \includegraphics[scale=0.8]{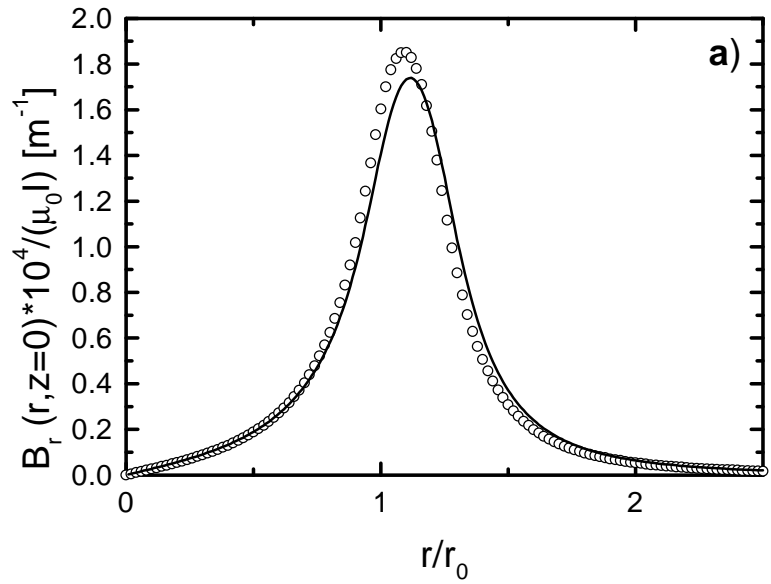}
   \includegraphics[scale=0.8]{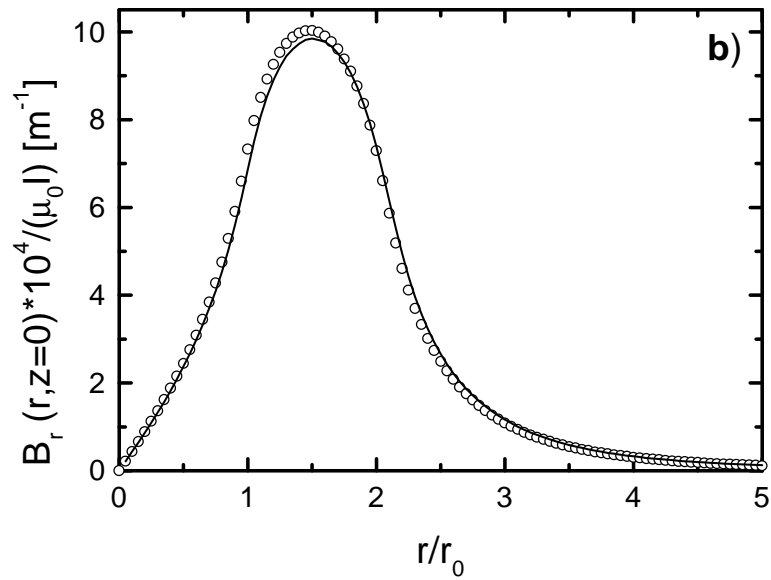}
  \caption{Radial component of the magnetic field for the two coils of cases \textit{a)} and \textit{b)}. The open circles and the lines represent the discrete and the continuous model respectively.}
  \label{fig3}
\end{figure}

Our general model has been checked in the particular case of a
single turn, single sheet coil by using the calculations of Dodd
and Deeds \cite{DODD}. In figure~\ref{fig2} we compare the results
of the calculations of the radial ($B_{r}$) and axial ($B_{z}$)
components of the induction field in the case of no sample for a
single turn coil with inner radius $r_{0}$=1 mm and wire diameter
equal to 50 $\mu m$. The fields have been calculated at a distance
$z=r_{0}$ from the coil surface ($z=0$). The two models give the
same result. In case of multi-turns coils, the continuous
approach\cite{MAWA1,DODD} has been widely used. We point out that
assuming the simple rectangular coil section of a continuous model
produces an over-estimation of the effective area covered by the
exciting coil current in respect to the discrete case, in which
the effective surface is the sum of N-turn discs of the diameter
of the coil wire. One would expect therefore an under-estimation
of the maximum value of the magnetic field compared to the
discrete case; the error decreases as the number of turns
increases approaching the continuous limit. To show this, we
compare the continuous \cite{MAWA1,DODD} and the discrete model
calculating the radial component of the magnetic field in two
cases: \textit{a)} a coil of 5 turns, 5 sheets; \textit{b)} a coil
of 22 turns, 24 sheets. The inner radius r$_{0}$ is 1 mm and the
sample to coil distance is h=0.1 mm; the wire diameter has been
assumed to be $50\mu m$. The results of these calculations are
reported in figure ~\ref{fig3}. As expected, the maximum B$_{r}$
field calculated with the continuous model is under-estimated
respect to the discrete one of 8 and 2\% for the case \textit{a)}
and \textit{b)} respectively. By using the methods proposed in
ref. \cite{Classen1,MAWA1}, this would imply an under-estimation
of the critical current density of the same amount.

\begin{figure}
\textwidth 9 cm
  \centering
  \includegraphics[scale=0.9]{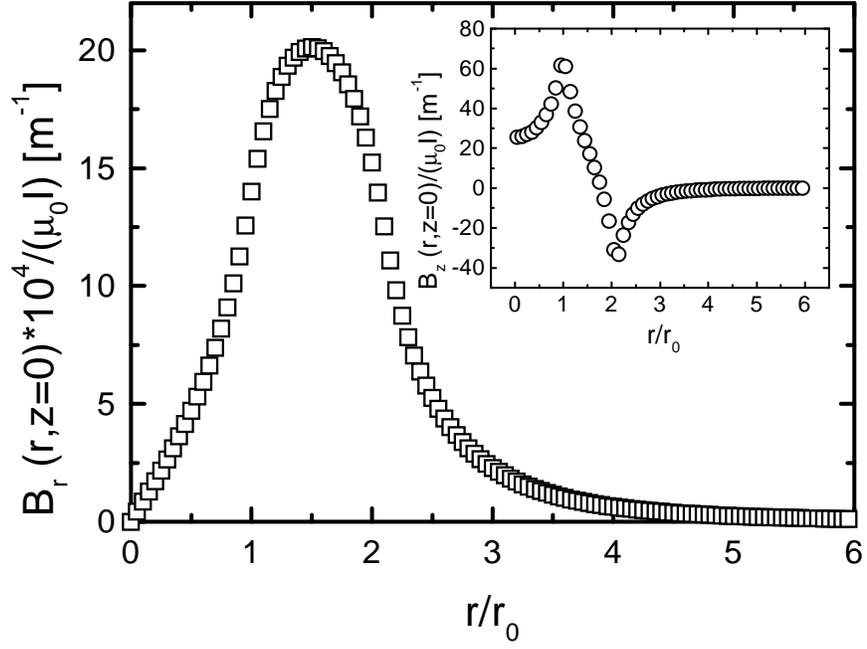}
  \caption{Radial component of the total induction field (open squares) for a multi-turn coil. In the inset the result for the normal component (open circles) is shown. Both components have been calculated at the sample surface as a function of r/r$_{0}$.}
  \label{fig4}
\end{figure}

Once the discrete model has been successfully tested, we have used
it to calculate the coefficient \textit{k(h)} in eq. ~\ref{0}.
Firstly, we have verified the condition to apply the Bean model
properly. To do so, we have calculated the total induction field
on the sample surface by using eq. ~\ref{8} and ~\ref{13}. The
parameters used in this calculation are: the sample thickness
$d$=700 nm, its magnetic penetration depth $\lambda=250$ nm for a
standard high quality YBCO sample \cite{LEMBE2,PRUSSEIT} at 77 K.
For the probe coil we have considered a 22 turns, 24 sheets coil
with the inner radius r$_{0}$= 1 mm and a wire diameter equal to
50 $\mu$m. The results of this calculation are reported in figure
~\ref{fig4}: the normal component B$_{z}$ is negligible in respect
to the radial component B$_{r}$. Moreover, it is worthwhile to
emphasize that the magnitude of B$_{r}$ is about two times larger
than in the case without sample shown in figure ~\ref{fig3}, in
the same point of the space, as we can expect within the framework
of a mirror image coil solution analysis. For the Bean model be
valid, the total induction field must be zero on the opposite side
of the film. This is true when the effective screening length
$\lambda_{screening}$ is equal to or less than the film thickness
$d$. From eq.~\ref{7}, in the case of a superconducting sample, it
results $\lambda_{screening}=\lambda \cdot coth(d/\lambda)$
~\cite{CHANG,GAUZZI}. When $\lambda_{screening} > d$ the error in
the estimation of J$_{C}$ may be very large: to estimate it
roughly, we can calculate the relative difference in the J$_{C}$
values when the film thickness equals $d$ and the effective
screening length $\lambda_{screening}$, i.e.
$|1-J_{C}(\lambda_{screening})/J_{C}(d)|$. For example, for a film
with $d$=200nm and $\lambda$=250nm, the error can rise up to 46\%.

\begin{figure}
\textwidth 9 cm
  \centering
  \includegraphics[scale=0.9]{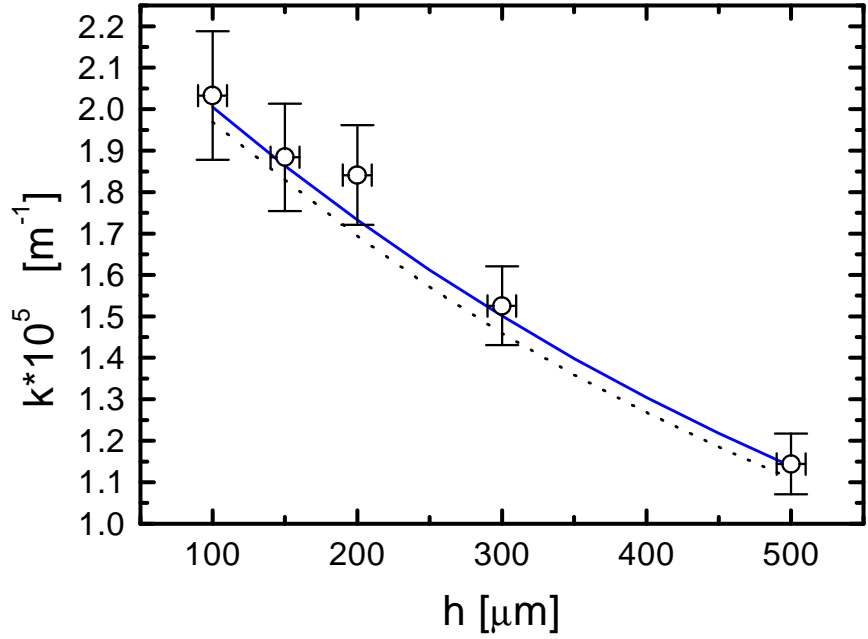}
  \caption{(Color Online) Determination of the scaling factor $k$
versus the sample to coil distance $h$. The continuous and the
dashed lines represent the behaviour expected for the discrete and
the continuous model respectively. The open circles represent the
$k$-values experimentally evaluated.}
  \label{fig5}
\end{figure}

To calculate the coefficient \textit{k} we take the maximum of the
coil function $f(r)$ plotted in figure ~\ref{fig3}. This
represents the spatial region where the field reaches its maximum
value. In the case of the copper coil reported above, our discrete
model gives the value of $k=2*10^{5} m^{-1}$ at a sample-coil
distance $h$=0.1 mm. Our calculation allows to estimate $k(h)$
with an error of 2\% for a minimum 10\% error in the determination
of $h$. $k(h)$ decreases as $h$ increases as it is shown in figure
~\ref{fig5}, because of the presence of terms $\propto
e^{-h\gamma}$ in the field expression. In the same graph, the
experimentally determinated values of $k$ at various distance are
also reported. The experimental values are in good agreement
within the error bars with the behavior predicted by the discrete
model. Details of the measurement will be given in the next
section. The exact knowledge of the relation linking \textit{k} to
\textit{h} makes this technique extremely flexible. For example,
in the case of superconducting tapes or large area films, the
probing coil must be moved all along the sample. Thus, to
correctly estimate J$_{C}$, it is necessary to know exactly the
magnetic coupling versus the sample to coil distance \textit{h}.

\section{Experimental}

To validate our discrete model we have inductively measured the
critical current density of a superconducting film of well known
transport properties. For the measure of the third harmonic signal
V$_{3}$ we used a Lock-in amplifier (Signal Recovery 7265). The
amplitude of the inducing sinusoidal current (the Lock-in
reference signal) circulating in the coil is measured by using a
high precision resistor in series with the coil. A commercial
highly linear amplifier increases the inducing sinusoidal signal.
The output signal is filtered by using a home made, low noise
notch filter to reduce the first harmonic component noise and to
avoid any saturation of the Lock-in. The temperature is monitored
by a temperature controller (Neocera2021) connected to a Cernox
resistor thermometer. The characteristics of the coil used in our
experiment have been reported in the previous paragraph.

\begin{figure}
\textwidth 9 cm
  \centering
  \includegraphics[scale=0.9]{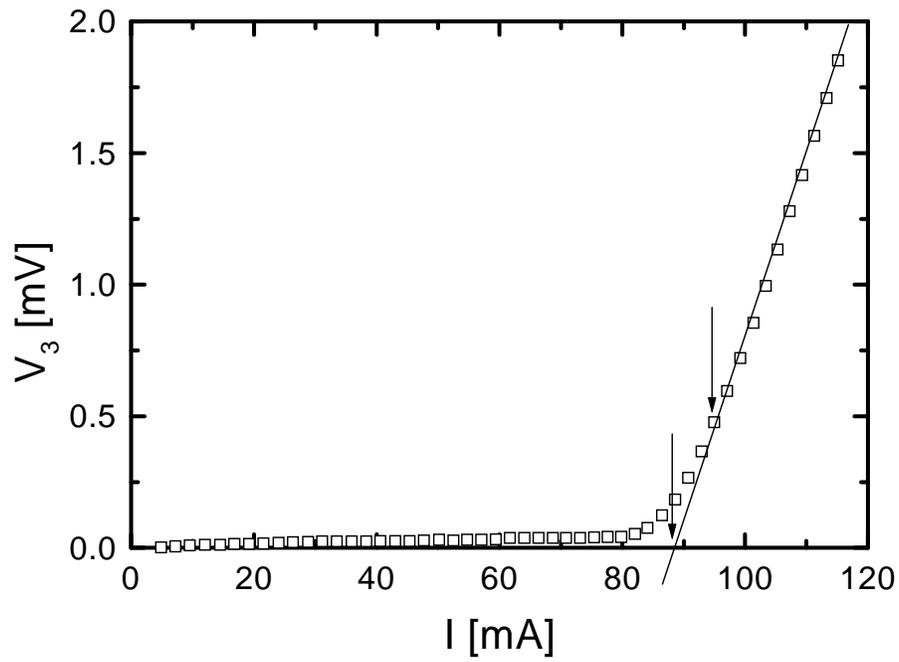}
  \caption{Determination of the critical current $I_{C0}$. The arrow indicates the last point below which the third harmonic voltage is no more linear as a function of I.}
  \label{fig6}
\end{figure}

The samples tested are 700 nm thick YBCO thin films produced by
Theva GmbH with dimensions 10 x 10 mm$^{2}$. The critical
temperature and the critical current density at liquid nitrogen
temperature given by the producer are 87.9 K and 2.57 MA/cm$^{2}$
(expressed in root mean square) respectively. The value of the
critical current $I_{C0}$ was experimentally determined looking at
the onset of non linearity by the extrapolation of the linear
portion in the V$_{3}$ versus I plot (see for example in figure
~\ref{fig6}~ and ref.~\cite{Classen1}). The value found for the
critical current density of the sample of fig.~\ref{fig6} is
$J_{C}=(2.6\pm0.3)MA/cm^{2}$ at 77 K, in very good agreement with
the data reported by the producer. For all the other samples
tested, $J_{C}$ has been found statistically well within the error
bars of the measurement presented above. The error on $I_{C0}$ is
estimated as the difference between the first point that is no
more aligned on the linear portion $V_{3}$ versus I and the
extrapolated $I_{C0}$ value (see the arrows on figure
~\ref{fig6}). This error is larger at low temperature because of
the increase of the critical current density: at higher $J_{C}$,
the noise level introduced by the amplifier is enhanced thus
producing a rounding in the lossy state transition.

We have repeated the same experiment at different sample-coil
spacings \textit{h}, by using a micrometric screw  with 10 $\mu m$
resolution placed under the sample holder, to calculate the
experimental scaling factor \textit{k} vs \textit{h} and comparing
it to the expected theoretical behaviour. We note some important
points: \textit{i)} the experimental values have been obtained by
considering the critical current density equal to the value given
by the producer; \textit{ii)} to minimize the error in the
determination of the sample-coil distance, we have set the zero of
the micrometric screw when the sample is in contact with the coil
($h=0\mu m$); \textit{iii)} the error bars for $k(h)$ have been
obtained by taking into account all the possible error sources
($d$ and $I_{C0}$) whereas for $h$ we have taken the minimum
resolution of the micrometric screw. As one can see in
figure~\ref{fig5}, there is a very good agreement between the
$k(h)$ values predicted by the discrete model and the experimental
ones.

\begin{figure}
\textwidth 9 cm
  \centering
  \includegraphics[scale=0.9]{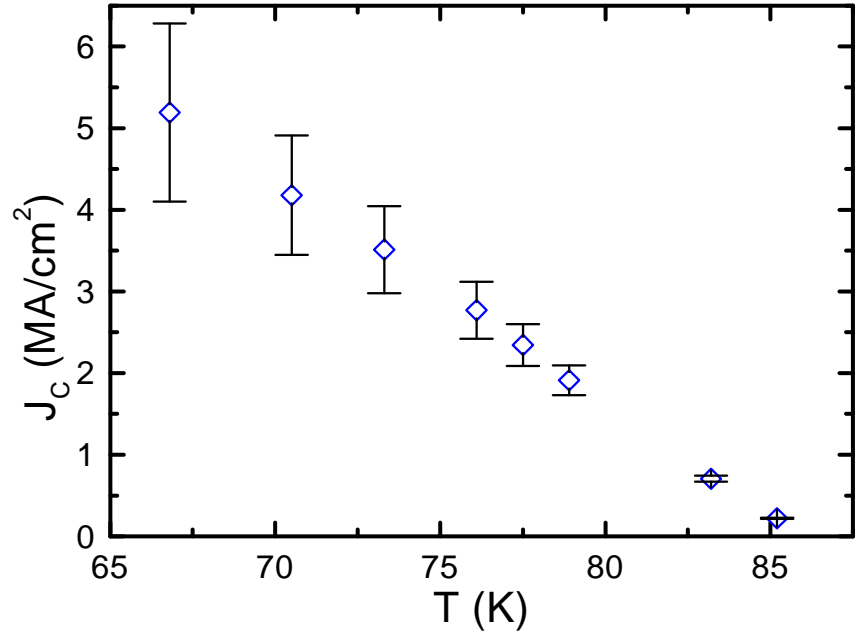}
  \caption{(Color Online) Critical current density J$_{C}$ versus temperature for a YBCO thin film.}
  \label{fig7}
\end{figure}

As a final test, we have plotted $J_{C}$ versus $T$ close to
$T_{c}$ in figure ~\ref{fig7}. The behaviour of the critical
current density is a linear function of the temperature. This is
usually explained as due to  the presence of Josephson coupling
between grains\cite{HALBR,MAWA4}. A more detailed analysis of the
$J_{C}(T)$ behaviour for $T>T_{C}/2$ goes beyond the scope of this
work and will be presented elsewhere.

\section{Conclusions}

In this paper we have revisited in detail an inductive and
contactless technique used for the measurement of the critical
current density in superconducting films. The inductive geometric
coupling factor has been calculated exactly by developing a
precise and realistic model describing the current circulating in
a real coil, and its image in a superconducting film placed in
close contact to it. In particular, the scaling law of the
geometric factor versus the sample-coil distance is found. This is
a fundamental issue to render this technique flexible, in order to
make possible a local non destructive evaluation of the critical
current density in large areas films or superconducting tapes just
moving a single probe coil at a known distance from the testing
sample. This discrete analysis has been successfully verified by a
comparison with similar continuous models presented in literature.
The model presented here is more realistic, relatively simple and
can describe the experimental data with a higher degree of
precision in particular in the case of coils with a low number of
turns contrarily to what happens using a continuous approach. The
model has been verified experimentally by measuring the critical
current density on YBCO films having well known transport
properties.

\begin{acknowledgments}

We thank Dr. M. Valentino for fruitful discussions. We are
grateful to Dr. V. De Luise, A. Maggio and S. Marrazzo for their
valuable technical assistance.\\
\\
\end{acknowledgments}

$^{*}$Corresponding author. E-mail address:
gianrico.lamura@na.infn.it.

\appendix
\section{The induction field expressions}

By using expressions (~\ref{5}-\ref{11}), (~\ref{13}-\ref{14}) and
the recursive relation for the derivative of the Bessel functions
we have calculated the radial and the normal induction field in
all the space with and without the sample. The expressions are the
following:

\textbf{without the sample:}

\begin{eqnarray}\label{16}
B_{r}(r,z)=\mu_{0}\frac{I_{0}}{2}\int^{\infty}_{0} \left[\sum_{n}
e^{-|z+h_{n}|\gamma} \sum_{m} r_{m} J_{1}(r_{m} \gamma)\right]
J_{1}(r\gamma) \gamma d\gamma
\end{eqnarray}

\begin{eqnarray}\label{17}
B_{z}(r,z)=\mu_{0}\frac{I_{0}}{2}\int^{\infty}_{0} \left[\sum_{n}
e^{-|z+h_{n}|\gamma} \sum_{m} r_{m} J_{1}(r_{m} \gamma)\right]
J_{0}(r\gamma) \gamma d\gamma;
\end{eqnarray}

One can define a geometric coil function\cite{MAWA1} $f(r)$ for
the radial component:

\begin{eqnarray}\label{23}
f(r)=\int^{\infty}_{0}  \left[\sum_{n} e^{-h_{n}\gamma} \sum_{m}
r_{m} J_{1}(r_{m} \gamma) \right] J_{1}(r \gamma) \gamma d \gamma.
\end{eqnarray}

\textbf{with the sample:}

z<0

\begin{eqnarray}\label{18}
B_{r}(r,z)=-\mu_{0}\frac{I_{0}}{2}\int^{\infty}_{0} \left[\sum_{n}
e^{-|z+h_{n}|\gamma} \sum_{m} r_{m} J_{1}(r_{m} \gamma)\right]
\left(1-F(\gamma,d,l)\right) J_{1}(r\gamma) \gamma d\gamma
\end{eqnarray}

\begin{eqnarray}\label{18}
B_{z}(r,z)=\mu_{0}\frac{I_{0}}{2}\int^{\infty}_{0} \left[\sum_{n}
e^{-|z+h_{n}|\gamma} \sum_{m} r_{m} J_{1}(r_{m} \gamma)\right]
\left(1-F(\gamma,d,l)\right) J_{0}(r\gamma) \gamma d\gamma
\end{eqnarray}

$0\leq z \leq d$

\begin{eqnarray}\label{19}
B_{r}(r,z)=\mu_{0}\frac{I_{0}}{2}\int^{\infty}_{0} \left[\sum_{n}
e^{-h_{n}\gamma} \sum_{m} r_{m} J_{1}(r_{m} \gamma)\right]
\frac{\partial G(\gamma,z,d,l)}{\partial z} J_{1}(r\gamma) d\gamma
\end{eqnarray}

\begin{eqnarray}\label{19}
B_{z}(r,z)=-\mu_{0}\frac{I_{0}}{2} \int^{\infty}_{0} \left[
\sum_{n} e^{-h_{n}\gamma} \sum_{m} r_{m} J_{1}(r_{m}
\gamma)\right] G(\gamma,z,d,l) J_{0}(r\gamma) \gamma d\gamma
\end{eqnarray}

\begin{eqnarray}\label{20}
\frac{\partial G(\gamma,z,d,l)}{\partial z}=\frac{ 2 \gamma
\left[\gamma l\sqrt{1+l^{2} \gamma^{2}}\cosh \left( \frac{z-d}{l}
\sqrt{1+l^{2} \gamma^{2}} \right) - \left( 1+l^{2}
\gamma^{2}\right) \sinh \left(\frac{z-d}{l} \sqrt{1+l^{2}
\gamma^{2}}
 \right) \right]} {(1+ 2 \gamma^{2}l^{2}) \sinh \left( \frac{d}{l}
\sqrt{1+l^{2}\gamma^{2}} \right)+ 2 \gamma l \sqrt{1+l^{2}
\gamma^{2}} \cosh \left(\frac{d}{l} \sqrt{1+l^{2} \gamma^{2}}
\right)}
\end{eqnarray}

z>d

\begin{eqnarray}\label{21}
B_{r}(r,z)=\mu_{0}\frac{I_{0}}{2}\int^{\infty}_{0} \left[\sum_{n}
e^{-h_{n}\gamma} \sum_{m} r_{m} J_{1}(r_{m} \gamma) \right] e^{-z
\gamma} J_{1}(r \gamma) \gamma d \gamma
\end{eqnarray}

\begin{eqnarray}\label{21}
B_{z}(r,z)=-\mu_{0}\frac{I_{0}}{2}\int^{\infty}_{0} \left[\sum_{n}
e^{-h_{n}\gamma} \sum_{m} r_{m} J_{1}(r_{m} \gamma) \right] \cdot
\left[1-e^{-z \gamma} -L(\gamma,l, d) e^{-d \gamma} \right]
J_{0}(r \gamma) \gamma d \gamma
\end{eqnarray}

where $F(\gamma,z,d,l)$,$G(\gamma,z,d,l)$ and $L(\gamma,d,l)$ have
been defined in eq.~\ref{7}, eq.~\ref{9} and eq.~\ref{11}
respectively.

Finally, it is possible to evaluate exactly the maximum electric
field induced on the surface of the sample (z=0). By using the
third Maxwell equation and eq.~\ref{8} we have:

\begin{eqnarray}\label{22}
E_{\vartheta}(r,z)=max \left\{ -\frac{\partial
A_{\vartheta}}{\partial t} \right\}=\mu_{0} I_{0} \omega  max
\left\{ f(r) \right\}
\end{eqnarray}

\bibliography{021524JAP}

\newpage

\end{document}